\pdfoutput=1
\documentclass[11pt]{article}

\usepackage[final]{acl}

\usepackage{times}
\usepackage{latexsym}
\usepackage{hyperref}
\usepackage{microtype}
\usepackage{graphicx}
\usepackage{subfigure}
\usepackage{booktabs} % for professional tables

\usepackage{listings}
\usepackage{xcolor}
\usepackage{float}
\usepackage{verbatim}

\usepackage{tcolorbox} 
\usepackage[inline]{enumitem}
\setlist[itemize]{leftmargin=4mm}
\setlist[enumerate]{leftmargin=4mm}
\newfloat{listing}{H}{lop}

\usepackage[T1]{fontenc}
\usepackage[utf8]{inputenc}

\usepackage{microtype}

\usepackage{inconsolata}

\usepackage{graphicx}

\usepackage{multirow}
\usepackage{makecell}

\title{A LLM-Driven Multi-Agent System for Professional Development of Mathematics Teachers}

\author{
 \textbf{Kaiqi Yang\textsuperscript{1}\thanks{Equal contribution.}},
 \textbf{Hang Li\textsuperscript{1}$^*$},
 \textbf{Yucheng Chu\textsuperscript{1}},
 \textbf{Ahreum Han\textsuperscript{2}},
 \\
 \textbf{Yasemin Copur-Gencturk\textsuperscript{2}},
 \textbf{Jiliang Tang\textsuperscript{1}}
 \textbf{Hui Liu\textsuperscript{1}}\thanks{Corresponding author.}
\\
 \textsuperscript{1}Michigan State University,
 \textsuperscript{2}University of South California
\\
 \texttt{\{kqyang,lihang4,chuyuch2,tangjili,liuhui7\}@msu.edu} \\
 \texttt{\{ahreumha,copurgen\}@usc.edu}
}

\begin{document}

\newcommand{\systemname}{\textsc{i-vip}}

\definecolor{myOrange}{rgb}{1,0.5,0}
\definecolor{myPurple}{HTML}{4045DB}
\definecolor{myLightPurple}{HTML}{D3C3F6}
\definecolor{myLightGray}{HTML}{F0F0F0}

\definecolor{myPurple}{HTML}{4045DB}\definecolor
{myLightYellow}{HTML}{FADA62}

\newcommand{\gbox}[3]{ 
   
    \begin{tcolorbox}[width=.999\columnwidth,  colback={#3}, halign=left] 
        \textbf{#1} 
        \tcblower
        {#2}
      \end{tcolorbox}
}

\newtcbox{\mcbox}[1][red]{on line,
arc=5pt,colback=#1!50!white,colframe=#1!90!black,
before upper={\rule[-3pt]{0pt}{10pt}},boxrule=1pt,
boxsep=0pt,left=6pt,right=6pt,top=2pt,bottom=2pt}

\newenvironment{metaverbatim}{\verbatim}{\endverbatim} 

\newcommand{\gboxx}[2]{ 
    \begin{tcolorbox}[width=.999\columnwidth,  colback={#2}, halign=left] 
        {#1}
    \end{tcolorbox}
}

\maketitle

\begin{abstract}
Professional development (PD) serves as the cornerstone for teacher tutors to grasp content knowledge. However, providing equitable and timely PD opportunities for teachers poses significant challenges. To address this issue, we introduce \systemname{} (Intelligent Virtual Interactive Program), an intelligent tutoring platform for teacher professional development, driven by large language models (LLMs) and supported by multi-agent frameworks. This platform offers a user-friendly conversational interface and allows users to employ a variety of interactive tools to facilitate question answering, knowledge comprehension, and reflective summarization while engaging in dialogue. To underpin the functionality of this platform, including knowledge expectation analysis, response scoring and classification, and feedback generation, the multi-agent frameworks are leveraged to enhance the accuracy of judgments and mitigate the issue of missing key points.

\end{abstract}
\section{Introduction}

Professional development (PD) is central to teacher tutoring, enhancing educators’ understanding of subject knowledge and critical thinking~\cite{copur2016sustainable,penuel2007makes}. Traditionally, PD has relied on methods such as in-person instruction or pre-scripted, rule-based computer systems, which are often rigid and unable to address diverse and large-scale user needs~\cite{glover2016investigating,burns2023barriers}. The advent of large language models (LLMs)~\cite{brown2020language,touvron2023llama} offers a promising alternative, providing equitable and convenient access to PD. Human users can interact conversationally with LLMs (e.g., ChatGPT) to clarify doubts and acquire new knowledge. However, PD frequently depends on domain-specific knowledge~\cite{fishman2013comparing}, such as pre-written materials containing questions, knowledge points, and expected answers, as well as media and interactive tools to support comprehension. Off-the-shelf LLM dialogue systems are designed for general use and cannot adequately meet these specialized needs. Moreover, PD tasks are often challenging, requiring LLMs not only to understand questions and answers but also to interpret users’ responses and follow-up inquiries.

To address these challenges, we leverage the capabilities of modern LLMs, particularly within multi-agent frameworks, to develop an effective and efficient solution for math teacher PD tutoring. We propose \underline{I}ntelligent \underline{V}irtual \underline{I}nteractive \underline{P}rogram, \systemname{}, a dialogue system designed for seamless use by educational domain experts. \systemname{} can integrate structured materials meticulously crafted by educators, automatically comprehend questions and knowledge points, provide users with timely conversational support, and present images and interactive tools to aid learning. \systemname{} is designed for mathematical tutoring of PD, especially the knowledge of proportion, ratio, and multiplication. To our knowledge, \systemname{} is the first teacher PD training system that fully harnesses LLM technology in combination with authentic educational content, serving as a case study for the application of LLMs and related technologies for societal good.

\section{Design Goals}

\begin{figure*}[!btph]
\vspace{-0.15cm}
    \centering
    \includegraphics[width=0.9\linewidth]{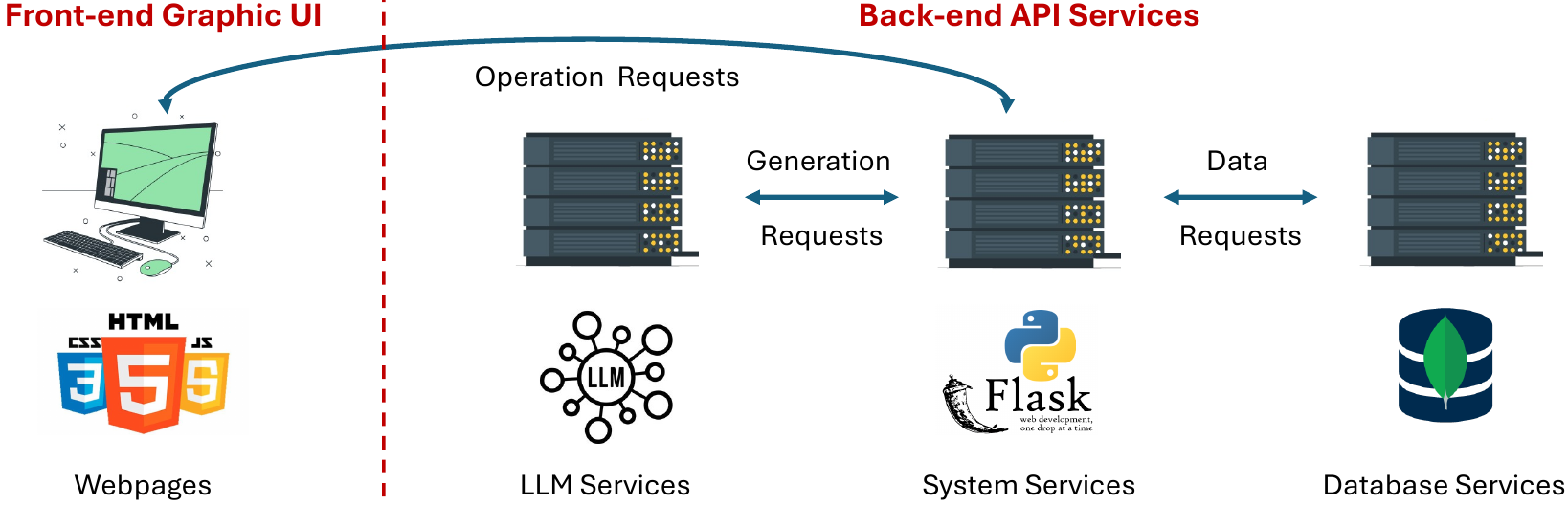}
    \caption{An illustration of the \systemname{}. Components of the system are communicated via the HTTP requests.}
    % \vspace{-.5cm}
    \label{fig:system_arch}
\vspace{-0.15cm}
\end{figure*}

The goal of our system is to provide a novel learning experience for learners engaged in professional development (PD) in pedagogical contexts. Unlike existing pre-programmed PD systems~\cite{cetin2016effects,lin2000using,gomaa2020ans2vec}, our system focuses on delivering a dynamic learning journey, where the learning content is adaptively generated rather than based on fixed templates. To enhance user engagement and emulate the experience of in-person learning, our system transforms interactions between users and the system into a conversational format. It employs open-ended questions instead of traditional multiple-choice formats, allowing users to pose inquiries and express their responses in more active and personalized ways. Additionally, our system is designed to integrate seamlessly with non-conversational tools, including graphical or programmatic applications. In the context of math PD tutoring, focusing on proportion and ratio, intuitive illustrations and visualizations are particularly beneficial for users. Data from users’ interactions with these tools is fused into the ongoing conversation between the user and the system, enriching the context for subsequent learning activities. Overall, we summarize the key design objectives of the system  as follows:

\paragraph{Adaptive Content:} The content provided during learning sessions is dynamically generated, taking into account both the user’s response history and the pre-defined learning syllabus. This ensures personalized and contextually relevant learning materials for each user.

\paragraph{Dialog-Based Interactions:} The primary mode of interaction is dialogue-based. The system poses open-ended questions and offers thought-provoking hints, encouraging users to respond actively. User responses are not expected to converge on a single \textit{correct} answer, and the system supports ongoing discussion and follow-up questions about the learning material.

\paragraph{Tool Compatibility:} The system is compatible with a variety of graphical or programmatic tools that enhance the learning experience beyond text-based communication. Users’ activity logs from these tools are incorporated as contextual information, enabling the system to generate relevant and tailored content throughout the learning process.

\paragraph{Feedback Mechanism:} The system is augmented with the feedback mechanism for users, particularly domain experts. Based on the collected feedback, developers can further refine the system's functionality and improve the user experience.
 
\section{System Design}
\label{sec:systemdesign}

\begin{figure*}[ht]
\vspace{-0.15cm}
    \centering
    \includegraphics[width=0.95\textwidth]{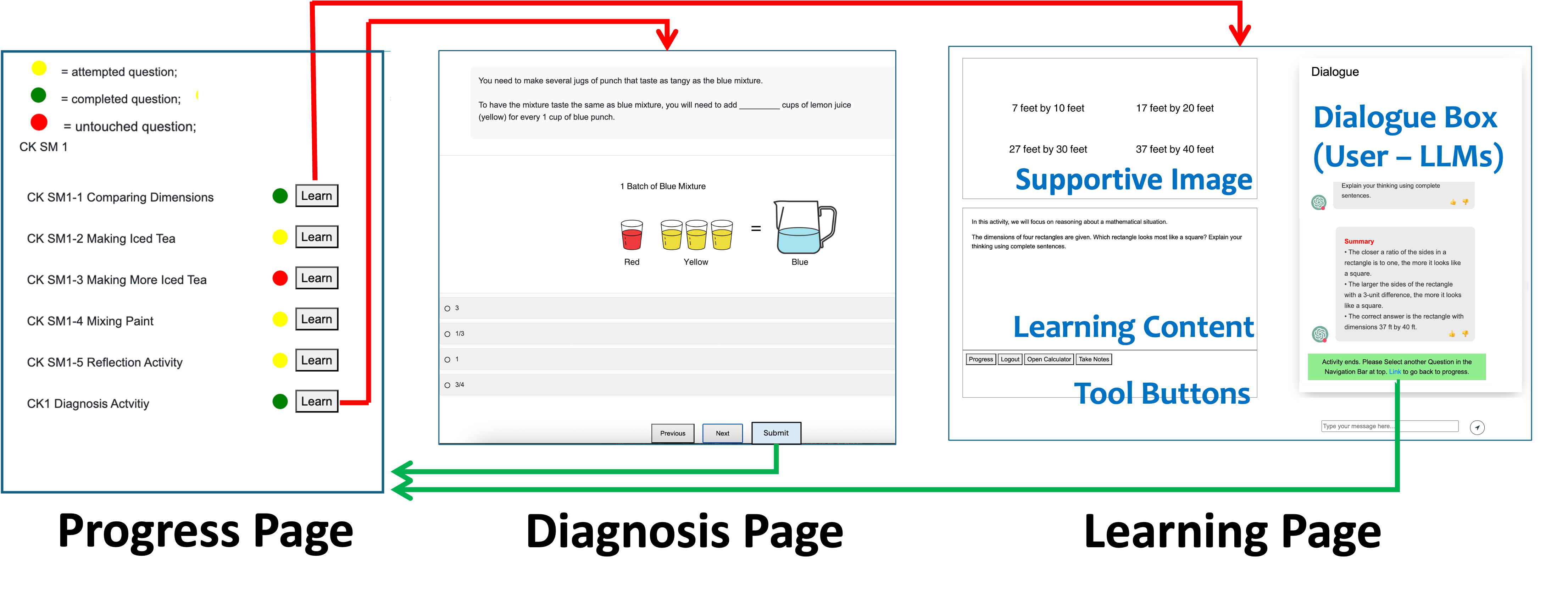}
    \vspace{-0.15cm}
    \caption{A demonstration of logistics of the system. Users can click the buttons (marked as \textit{Learn}) in the \textit{progress page} (left) and redirect to corresponding \textit{learning page} (right) or \textit{diagnosis page} (middle).}
    % \vspace{-.5cm}
    \label{fig:flow}
\vspace{-0.25cm}
\end{figure*}

Our system, \systemname{}, is an online interactive professional development (PD) platform for mathematic learning practice, implemented through a combination of front-end graphical user interfaces (UI) and back-end API services. The system is built using the Flask framework~\cite{grinberg2018flask}, and communication between the front-end UI and the back-end services is facilitated via HTTP requests. To enable different functionalities in the front-end UI, we implement various JavaScript functions linked to specific UI elements. These functions are responsible for collecting diverse types of user interaction data and sending these data to the corresponding back-end services via HTTP requests. The outputs from these services, including newly generated response content for users and system operation commands, are received and handled by the JavaScript functions to update the UI accordingly. On the back end, services are implemented through a combination of local Python functions and remote API calls, supporting database queries and large language model (LLM) generation tasks. The database stores all relevant materials for the PD courses and logs each user’s interaction history over time. The LLM generation component is a flexible module integrated with various techniques, such as in-context learning~\cite{dong2022survey} and retrieval-augmented generation~\cite{gao2023retrieval}, enabling dynamic generation of system responses within learning dialogues. An overview of the system is illustrated in Figure~\ref{fig:system_arch}. In the following sections, we present the details of each component in our system.

\subsection{User Interfaces}
Our user interface consists of three main components: the \textit{progress pages}, \textit{learning pages}, and \textit{diagnosis pages}. We organize the logic and navigation of these pages as illustrated in Figure~\ref{fig:flow}, following this workflow: (1) After logging in, users are directed to the \textit{progress page}, where a list of available activities is displayed. (2) Users can choose to enter either a \textit{learning page} or a \textit{diagnosis page} by clicking the \textit{Enter} button located on the right-hand side of each activity listing. (3) Upon completing an activity, users are redirected back to the \textit{progress page} to select their next activity.

\subsubsection{Progress Page}
The UI of the \textit{progress page} (left section of Figure~\ref{fig:flow}) displays a list of activities available on the platform, along with the corresponding modules and learning progress. Overall, we organize the page in a hierarchical structure: CK/PCK → Module → Activity. Specifically, CK (Content Knowledge) and PCK (Pedagogical Content Knowledge) are the broad knowledge domains under which various modules are organized. Each module (e.g., CKSM1) groups related learning content and objectives, focusing on one topic in mathematics, such as proportional relationship, covariance and invariant ratios, etc. Each activity (e.g., CKSM1-1, CKSM1-2) represents a specific learning task or problem within a module, such as the fixed difference between variables. In our current system, we have prepared 8 modules and 51 activities in total. For each mathematical activity, the progress page displays the problem title and the user’s learning status, which can be one of the following: \textit{not attempted}, \textit{attempted but not completed}, or \textit{completed}. The platform records each user’s progress across different activities, allowing users to seamlessly resume learning from where they left off. At the end of each module, we include a diagnosis activity designed to assess learning outcomes. These diagnosis activities are initially locked and become accessible only after the user has completed all activities within the corresponding module. This mechanism reviews the math PD knowledge in corresponding module and enables users to confirm their mastery of the module by taking the test. To enter any activity listed on the progress page, users can click the \textit{Enter} button. Upon clicking, the platform redirects them to the appropriate \textit{learning page} or \textit{diagnosis page} corresponding to the selected activity.

\subsubsection{Learning Page}

The \textit{learning page} displays the selected PD learning module contents, including the contents of math questions, images (optional), dialogue box, and buttons for interactive tools (the right sub-figure of Figure~\ref{fig:flow}). The questions and corresponding images are designed by domain experts and displayed in the left section of \systemname{}. In the dialogue box on right side, hints (including rubrics, questions and guidance) are provided to encourage users to respond. After the user replies, the information is sent to the backend (see Section \ref{sec:backend}) and the feedback is displayed promptly. There are two forms of feedback: text and tools. Text feedback includes answers, instructions, and questions to assist users in addressing the problems. Tools include interactive tools and image displays: the platform shows relevant interactive tools (see Section \ref{sec:tools}) to help users understand mathematical concepts; the other type of tool is image display, where the platform presents images designed by domain experts to provide hints or examples in response to user answers. Considering the knowledge of math PD, interactive tools and images are more efficient to present the math concepts and relationships. When the users fully grasp all knowledge points, the page displays a review and summary of the problem and users are prompted to exit the current mathematical activity to the progress page and start to work on the next activity.

\subsubsection{Diagnosis Page}

As mentioned before, the \textit{diagnosis page} is unlocked after all mathematical activities of a module are completed. Upon entering the diagnosis page, a series of test questions, mostly multiple-choice and reviewed by domain experts, are displayed in sequence. We record user answers to support their professional development (PD) and potential future studies. The collected data, which includes selected answers and corresponding timestamps, is structured as a time series. For multiple-choice questions with multiple answers, each option is recorded individually. Additionally, since users can revisit and revise their answers to previous questions, the history of attempts is also captured. This comprehensive dataset from the diagnosis module provides valuable insights for educational researchers to analyze learning progress. An example of the diagnosis page is shown as the middle sub-figure of Figure \ref{fig:flow}.

\subsubsection{Interactive Tools~\label{sec:tools}}

\begin{figure}[ht]
\vspace{-0.15cm}
    \centering
    \includegraphics[width=0.75\columnwidth]{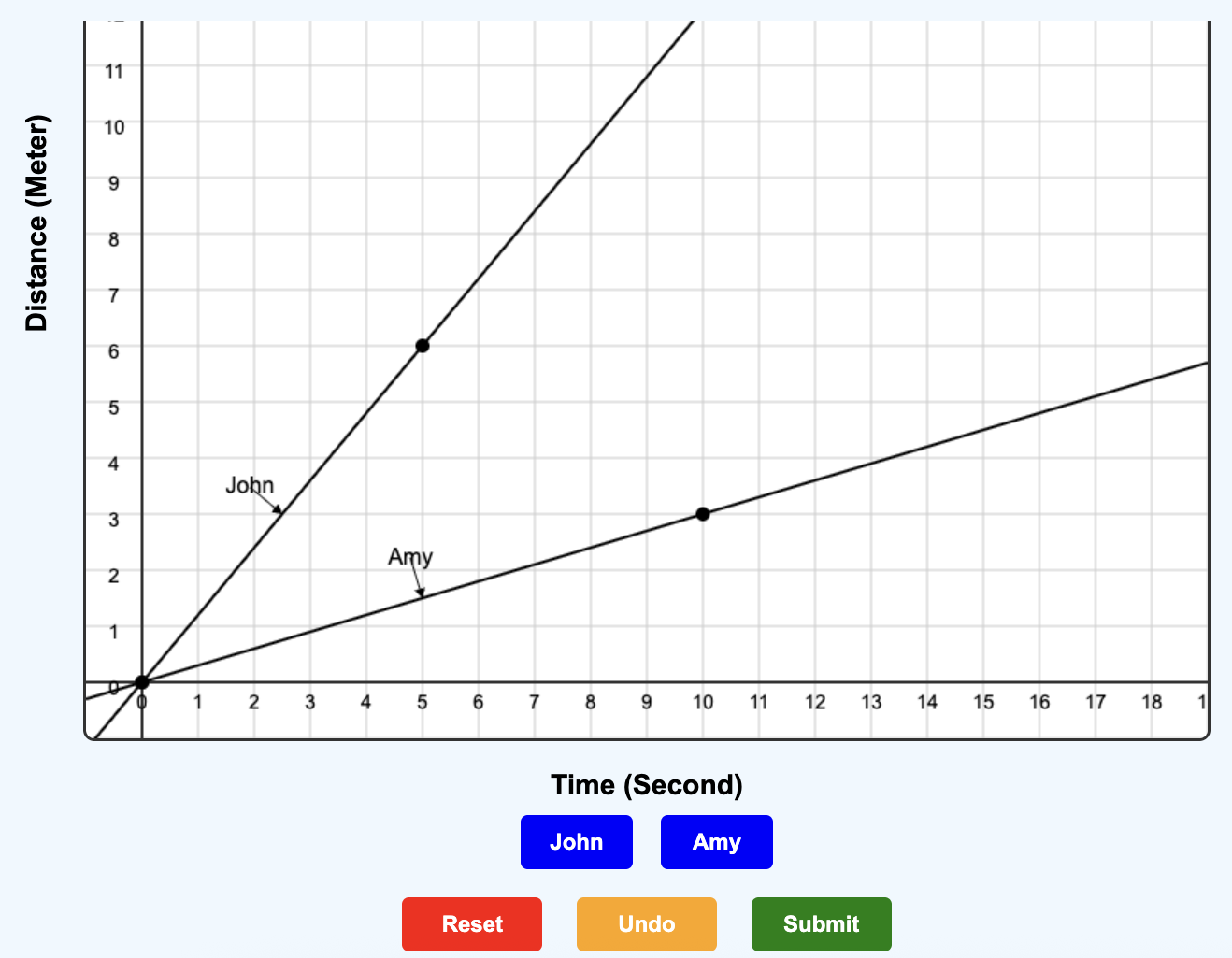}
    \caption{Plot of draw board (two-line). Users can draw graphs for linear functions and comprehend the mathematical concepts.}
    \label{fig:toolT}
\vspace{-0.15cm}
\end{figure}

\begin{figure}[ht]
\vspace{-0.15cm}
    \centering
    \includegraphics[width=0.8\columnwidth]{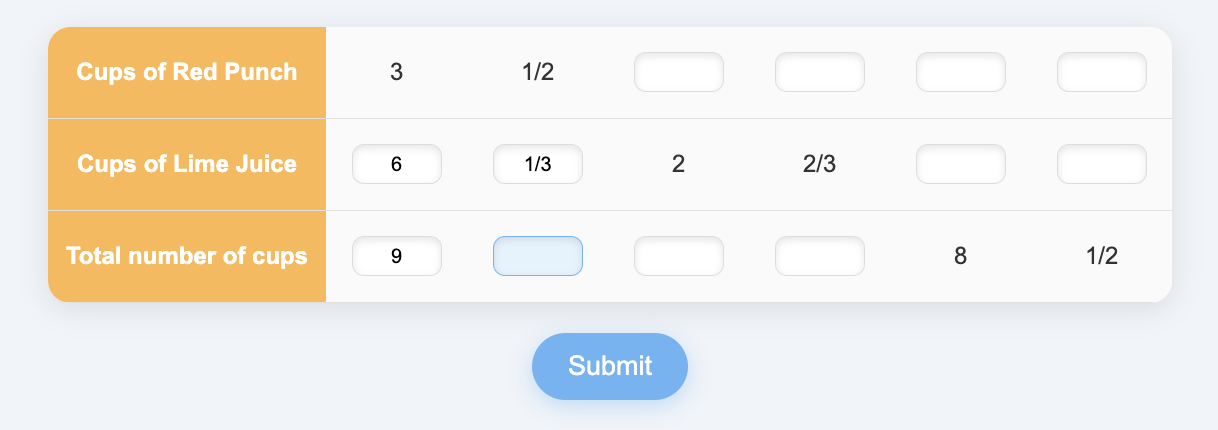}
    \caption{Plot of draw board (fill-table). Users can input numbers into the table to explore the invariant ratios between variables despite varying magnitudes. }
    \label{fig:toolTable}
\vspace{-0.15cm}
\end{figure}

Besides the learning page and diagnosis page, we design several interactive tools to facilitate the mathematical professional development, which are presented as pop-up web pages. The tools can be automatically displayed during the learning process, or called by users clicking the corresponding buttons. Examples of interactive tools include notebooks and draw boards. Notebook (as shown in Figure \ref{fig:toolN} in Appendix~\ref{sec:app-tool}) is designed to take notes of user' insights or key points; the content of a notebook is shared across different learning stages, thus allowing users to review their previous notes at any time. Draw-boards are a group of interactive tools with adjustable graphics, which are suitable to present the mathematical concepts and values in proposed questions. For instance, the two-line board (as shown in Figure \ref{fig:toolT}) enables users to draw multiple linear function graphs to comprehend the relationship between slope and velocity, and the fill-table board (as shown in Figure \ref{fig:toolTable}) allows users to fill in a table with numbers and hints, thereby understanding the relationship between ratio and value. 

\subsection{Backend Services~\label{sec:backend}}
\begin{figure}[ht]
\vspace{-0.25cm}
    \centering
    \includegraphics[width=0.9\columnwidth]{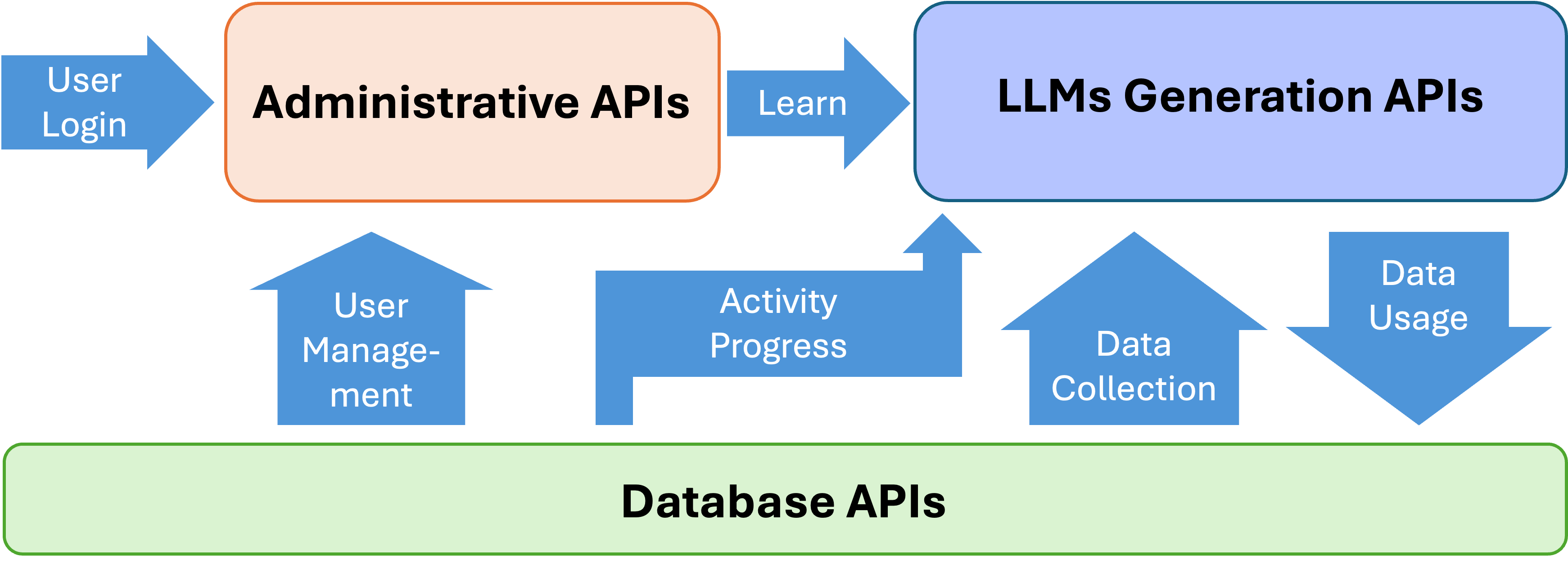}
    \caption{Plot of Backend Services.}
    \label{fig:backend}
\vspace{-0.25cm}
\end{figure}

Backend Services utilize APIs to communicate with the user interface and the functions on server, driving the interactions between human users and the computer-based modules. As shown in Figure~\ref{fig:backend}, the backend APIs can be categorized into three types. \textit{LLMs Generation} is the core of system, which calling APIs to generating responses for dialogues; \textit{Database Construction and Querying} consists of APIs taking records of user and system behaviors, as well as retrieving the relevant behavior history to support information processing; \textit{Administrative Management} covers APIs for platform and activity workflows, including the learning progress of users (accounts) and the corresponding mathematical activities.

\subsubsection{LLMs Generation} 
\begin{figure}[ht]
    \centering
    \includegraphics[width=0.85\columnwidth]{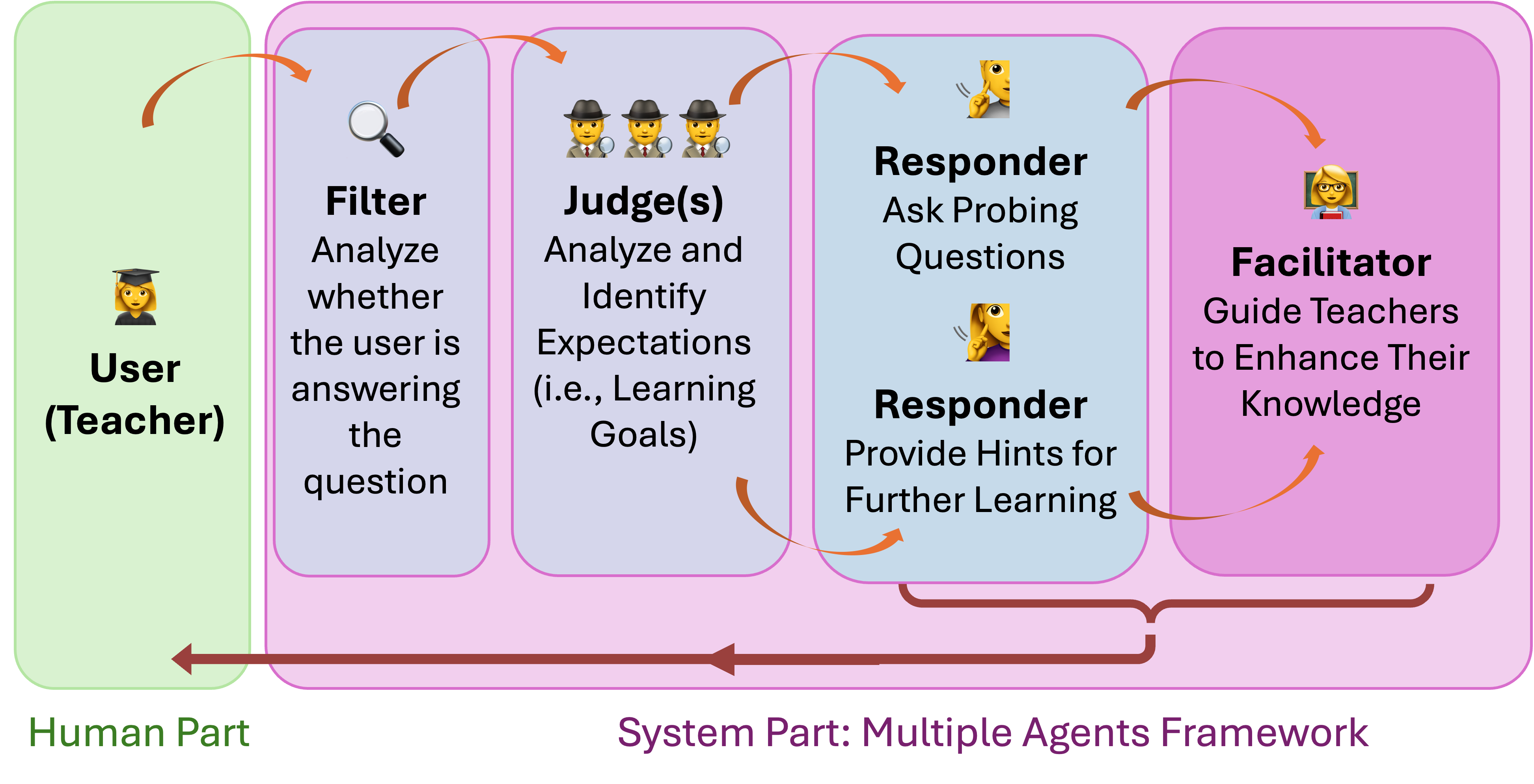}
    \caption{Multi-agent framework of LLMs generation for a life-cycle of dialogue rounds.}
    \vspace{-0.25cm}
    \label{fig:multiagent}
\end{figure}

After the user responds, the content of the current activity (text and the results of interactive tool usage) is sent to the backend LLMs, and LLMs are asked to generate proper response message or behaviors. We design multi-agent frameworks for \systemname{}, as shown in Figure~\ref{fig:multiagent}, which processes the user messages by detecting intends, correctness, grasp of knowledge, etc. and generating proper responses to proceed with the dialogue. 

The multi-agent framework, as illustrated by Figure~\ref{fig:multiagent}, consists of several components for different tasks. Given a message from human user (\textit{Teacher}): \textit{Filter} detects the intend of user and navigate the following procedures; \textit{Judge(s)} are agents checking the grasp of knowledge and missing points; \textit{Responder(s)} are agents generating tailored responses, e.g. answering user's questions, providing hints for future learning, etc.; \textit{Facilitator} indicates agents managing the dialogue and sending message back to the user. With joint efforts of all agents, \systemname{} tracks the progress of PD tutoring and make timely interactions with users.

For tasks such as expectation judgment, multiple LLM-agents are maintained to cooperate because this task requires checking the user's response and the mathematical problem, as well as the corresponding multiple expectations (knowledge points), and reporting which expectations the user's response correctly covers. Since this task is challenging, a single call to LLMs may generate inaccurate judgments, such as ignoring some key points or aspects, which are more likely to be mentioned when multiple agents work together, thus generating more comprehensive and accurate judgments.

\subsubsection{Database Construction and Querying}
A database is connected to support the \systemname{}. User and system behaviors, including text dialogues and the use of interactive tools,  are recorded and queried when needed. For data collection, we keep records of all trails of behaviors, including the text messages sent by users, their actions in using the interactive tools, as well as the responses of system, judgment of users' answer to interactive tools, and the supportive materials. Besides, all behaviors are kept with the corresponding timestamps. For data usage, we dynamically query the history of user and system behaviors for LLMs inference, such as the prior rounds of dialogue between them to check whether the user fully understand the knowledge.

\subsubsection{Administrative Management}

These APIs are designed to support various platform workflow management functions. Specifically, the system operates with two major workflows: (1) the platform-wide learning process workflow, and (2) the activity-specific learning workflow. For the platform-wide workflow, the API services handle operations such as user identity registration and verification, and the recording and tracking of learning progress. By responding to requests, such as displaying the status of activities on the progress page, these APIs guide users through the learning process at the platform level. For the activity-specific workflow, the APIs take the output responses from the LLM generation services as input and determine the system’s next adaptive response within the dialogue with the user. More specifically, because learning activities in our system typically divide the learning process into multiple stages, with several expectations to be met at each stage, the activity-specific APIs analyze the user’s latest responses to assess progress and decide on the next appropriate system action. This may include providing hints to help users discover current learning expectations, moving on to the next unmet expectation, or even skipping to subsequent stages. Through the control operations provided by these APIs, the system can adaptively organize its dialogue and interactions with different users. Finally, the system also offers external data access via dedicated sockets, allowing developers and domain experts to review users’ learning progress and access logs of question-response histories. This functionality supports external system monitoring and facilitates further academic research.

\section{Evaluation and Improvement}

To evaluate the validity of \systemname{} system, we conducted an online user study involving five Ph.D. and postdoctoral students from the university’s education department. The participants were asked to share their impressions of the system through a structured usage experiment. Specifically, each evaluator was instructed to complete all 51 available activities and indicate their satisfaction with the system’s responses by using the upvote or downvote buttons displayed at the bottom of each chat bubble (we present the interface of feedback in Appendix~\ref{sec:app-tool}). By collecting and analyzing these feedback clicks, we observed an overall satisfaction rate of 97.49\% for the system’s responses. The detailed statistics regarding user feedback for responses generated by different system components are presented in Table~\ref{tab:feedback}.
 
\begin{table}[!btph]
\vspace{-0.15cm}
\centering
\caption{Detailed statistics on the positive and negative feedback provided by human testers for each component of the system.}
\label{tab:feedback}
\vspace{-0.15cm}
\resizebox{0.48\textwidth}{!}{
\begin{tabular}{@{}c|cccc@{}}
\toprule
\textbf{Component} & \textbf{\# Response} & \textbf{\# Positive} & \textbf{\# Negative} & \textbf{\% Positive} \\ \midrule
Filter & 1,900 & 1,791 & 109 & 94.26 \\
Judger & 1,538 & 1,419 & 119 & 92.26 \\
Responder & 7,135 & 6,999 & 136 & 98.09 \\
Facilitator & 1,538 & 1,538 & 0 & 100.0 \\
Tools & 2,380 & 2,380 & 0 & 100.0 \\ \midrule
Total & 14,491 & 14,127 & 364 & 97.49 \\ \bottomrule
\end{tabular}}
\vspace{-0.15cm}
\end{table}

In addition, to demonstrate the effectiveness of the feedback mechanism in improving system performance, we conducted a complementary offline evaluation. In this experiment, we leveraged automatic prompt optimization~\cite{chu2024llm} and in-context learning~\cite{li2024knowledge} techniques to address the failure cases identified during the online test. For this evaluation, we divided the failure cases into five equal splits and performed 5-fold cross-validation to assess the impact of the improvement methods. For each k-th fold, we used the remaining four folds as training data to refine the existing prompts and as in-context examples. We then calculated the accuracy of the system on the k-th fold. The accuracy results of the improved models across the 5-fold cross-validation experiment are reported in Table~\ref{tab:improve}.

\begin{table}[!btph]
\vspace{-0.25cm}
\centering
\caption{Accuracy improvement with different techniques, \textit{Rubric-Opt}~\cite{chu2024llm} and \textit{Few-Shot}~\cite{li2024knowledge} are methods with improved prompts or in-context learning data, while \textit{Both} combines these two methods. All values are improvement and the baseline performance is 0, because the dataset consists of failure cases of the system before update.}
% \vspace{-0.15cm}
\label{tab:improve}
\resizebox{0.38\textwidth}{!}{
\begin{tabular}{@{}c|ccc@{}}
\toprule
\ \ \ \ \textbf{Method}\ \ \ \ & \textbf{Rubric-Opt} & \textbf{Few-Shot} &\ \ \ \ \textbf{Both}\ \ \ \ \\ \midrule
Improve & 34.83 & 73.03 & 74.16 \\ \bottomrule
\end{tabular}}
\vspace{-0.25cm}
\end{table}

\section{Future Directions}
\systemname{} is capable of real-time interaction with users, conducting professional development training through dialogue and interactive tools. Meanwhile, we have also identified the following potential directions for optimization.
\begin{itemize}

\item \textbf{Personalized Feedback}: \systemname{} generates responses based on current round of interactions without considering user-specific information. Future work will explore adapting to diverse user knowledge levels (mathematical proficiency, language ability, cognitive capacity) and providing tailored feedback that better captures user intentions.

\item \textbf{Long-Distance Context}: \systemname{} currently relies on short-term context (current question-answer pair, recent dialogue rounds, learning progress). To enhance math professional development, we aim to efficiently incorporate longer dialogue history, previously mastered knowledge, and track long-term user focuses and intentions.

\item \textbf{Balancing Accuracy and Efficiency of Multi-Agent Frameworks}: While the multi-agent framework improves accuracy, it incurs higher costs and time overhead. Therefore, we will consider: How to efficiently invoke the multi-agent framework? How to automatically decide whether to invoke the multi-agent framework and the scale of agents for different tasks?
\end{itemize}

\section{Conclusion}
We propose \systemname{}, an interactive dialogue platform powered by Large Language Models (LLMs) and supported by multi-agent frameworks. The core components include tutoring activities designed by domain experts, interactive learning tools, the expectation judgment module, the feedback generation module, and reflective modules. Enabled by the robust capabilities of LLMs and the collaborative dynamics of multi-agent frameworks, \systemname{} can promptly detect and respond to the progress and intentions of human users, generating appropriate responses that significantly enhance teachers' professional development. \systemname{} offers inclusive and cost-effective opportunities for the PD of mathematics teachers. Future directions include providing personalized feedback tailored to the diverse knowledge backgrounds and personalities of human users, and efficiently leveraging larger-scale contextual information and dialogue history.

\bibliography{paper}

\appendix
\begin{figure*}[ht!]
    \centering
    \includegraphics[width=0.8\textwidth]{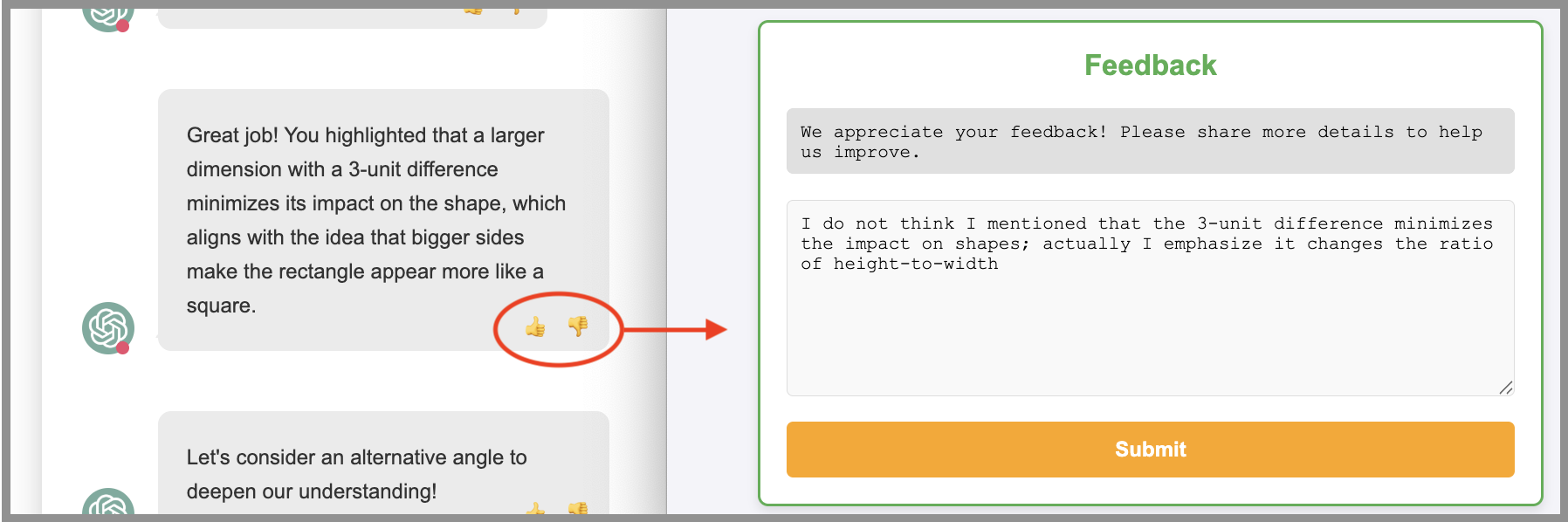}
    \caption{Plot of the feedback tool. Users can click the thumb buttons to indicate their judgment to the system responses, and it's optional to state their suggestions or opinions for future system updates.}
    \label{fig:feedback}
\end{figure*}

\section{Illustrations of Interactive and Feedback Tools}
\label{sec:app-tool}

\begin{figure}[ht]
    \centering
    \includegraphics[width=0.9\columnwidth]{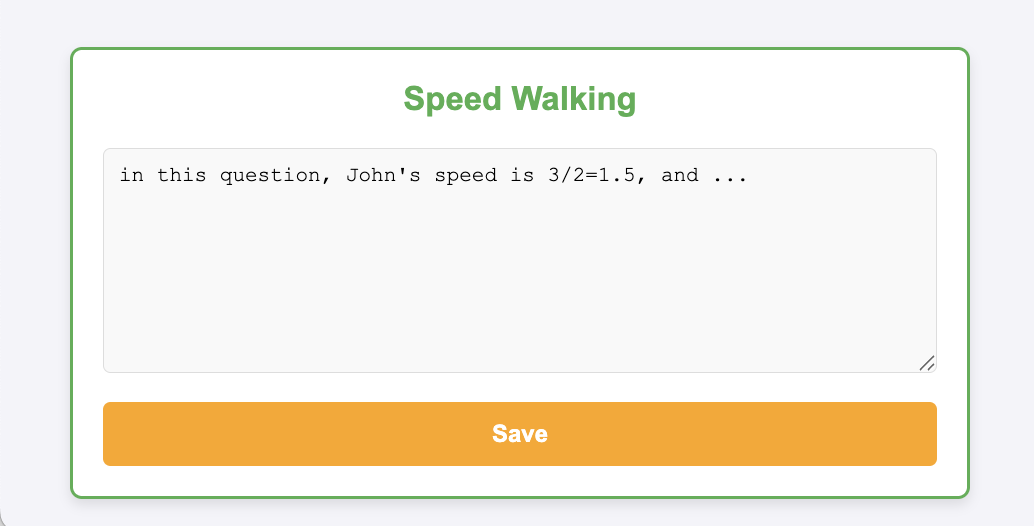}
    \caption{Plot of the notebook tool. Users can take notes and review them when learning. }
    \label{fig:toolN}
\end{figure}

In this section, we presents the figures of interactive tools as introduced in Section~\ref{sec:tools}, as well as the feedback tool. Figure~\ref{fig:toolN} is the notebook, which allows users to take notes and review them when learning. 

In the dialogue box, we append a pair of upvote and downvote buttons for all responses generated by the system. As shown in Figure~\ref{fig:feedback}, the users can click the buttons to judge the quality and correctness of system responses, and a note is optional to further explain their suggestions. This is particularly helpful when the domain experts provide insights to guide the developers improve the system.

\end{document}